\documentclass[twocolumn]{aastex61}

\received{October 17, 2017}
\revised{December 3, 2017}
\accepted{December 5, 2017}
\submitjournal{ApJ}

\shorttitle{Particle Acceleration at the Inward Shock in Cassiopeia A}
\shortauthors{Sato et al.}
\begin{document}

\title{X-ray Measurements of the Particle Acceleration Properties at Inward Shocks in Cassiopeia A}

\correspondingauthor{Toshiki Sato}
\email{toshiki@astro.isas.jaxa.jp}

\author{Toshiki Sato}
\affiliation{Department of Physics, Tokyo Metropolitan University, 1-1 Minami-Osawa, Hachioji, Tokyo 192-0397}
\affiliation{Department of High Energy Astrophysics, Institute of Space and Astronautical Science (ISAS), Japan Aerospace Exploration Agency (JAXA), 3-1-1 Yoshinodai, Sagamihara, 229-8510, Japan}

\author{Satoru Katsuda}
\affiliation{Department of Physics, Faculty of Science \& Engineering, Chuo University, 1-13-27 Kasuga, Bunkyo, Tokyo 112-8551, Japan}

\author{Mikio Morii}
\affiliation{The Institute of Statistical Mathematics, 10-3 Midori-cho, Tachikawa, Tokyo 190-8562, Japan}

\author{Aya Bamba}
\affiliation{Department of Physics, The University of Tokyo, 7-3-1 Hongo, Bunkyo-ku, Tokyo 113-0033, Japan}
\affiliation{Research Center for the Early Universe, School of Science, The University of Tokyo, 7-3-1 Hongo, Bunkyo-ku, Tokyo 113-0033, Japan}

\author{John P. Hughes}
\affiliation{Department of Physics and Astronomy, Rutgers University, 136 Frelinghuysen Road, Piscataway, NJ 08854-8019, USA}
\affiliation{Center for Computational Astrophysics, Flatiron Institute, 162 Fifth Avenue, New York, NY 10010, USA}

\author{Yoshitomo Maeda}
\affiliation{Department of High Energy Astrophysics, Institute of Space and Astronautical Science (ISAS), Japan Aerospace Exploration Agency (JAXA), 3-1-1 Yoshinodai, Sagamihara, 229-8510, Japan}

\author{Manabu Ishida}
\affiliation{Department of High Energy Astrophysics, Institute of Space and Astronautical Science (ISAS), Japan Aerospace Exploration Agency (JAXA), 3-1-1 Yoshinodai, Sagamihara, 229-8510, Japan}

\author{Federico Fraschetti}
\affiliation{Department of Planetary Sciences and Astronomy, University of Arizona, Tucson, AZ, 85721, USA}

%\author{\bf Note. the order will be adjusted}

\begin{abstract}
We present new evidence that the bright non-thermal X-ray emission
features in the interior of the Cassiopeia A supernova remnant (SNR)
are caused by inward moving shocks based on {\it Chandra} and {\it
  NuSTAR} observations. Several bright inward-moving filaments were
identified using monitoring data taken by {\it Chandra} in
2000--2014. These inward-moving shock locations are nearly coincident
with hard X-ray (15--40 keV) hot spots seen by {\it NuSTAR}. From
proper motion measurements, the transverse velocities were estimated
to be in the range $\sim$2,100--3,800 km s$^{-1}$ for a distance of
3.4 kpc. The shock velocities in the frame of the expanding ejecta
reach values of $\sim$5,100--8,700 km s$^{-1}$, slightly higher than
the typical speed of the forward shock.  Additionally, we find flux
variations (both increasing and decreasing) on timescales of a few
years in some of the inward-moving shock filaments. The rapid
variability timescales are consistent with an amplified magnetic field
of $B \sim$ 0.5--1 mG.  The high speed and low photon cut-off energy
of the inward-moving shocks are shown to imply a particle diffusion
coefficient that departs from the Bohm regime ($k_0 = D_0/D_{\rm
  0,Bohm} \sim$ 3--8) for the few simple physical configurations we
consider in this study. The maximum electron energy at these shocks is
estimated to be $\sim$8--11 TeV, smaller than the values of
$\sim$15--34 TeV inferred for the forward shock.  Cassiopeia A is
dynamically too young for its reverse shock to appear to be moving
inward in the observer frame.  We propose instead that the
inward-moving shocks are a consequence of the forward shock
encountering a density jump of $\gtrsim$ 5--8 in the surrounding
material.
\end{abstract}


\section{Introduction} \label{sec:intro}
Shock waves in young supernova remnants (SNRs) are excellent
astrophysical laboratories for understanding the process of particle
acceleration. An important observational advance was the imaging of
synchrotron X-ray--emitting rims in SN1006 \citep{1995Natur.378..255K}
which strengthened arguments for the shock acceleration of electrons
to TeV energies in the forward shocks of young SNRs.  More recently,
GeV and TeV $\gamma$-ray spectral observations have been shown to be
well described by the decay of pions produced by energetic protons
interacting with ambient gas \citep[e.g.,][]{2013Sci...339..807A},
providing strong evidence for proton acceleration in SNRs.  From these
studies as well as several others using a variety of different
techniques, it is now widely accepted that electron and proton
acceleration occurs in SNRs.  Diffusive shock acceleration
\citep*[DSA:][]{1978MNRAS.182..443B,1987PhR...154....1B} in the
forward shocks of remnants is the leading candidate for the process
and can account for the majority of Galactic cosmic rays.  On the
other hand, observational evidence for particle acceleration at SNR
reverse shocks, which propogate inward through the expanding ejecta,
has not been clearly demonstrated, although there have been some
theoretical studies
\citep[e.g.,][]{2005A&A...429..569E,2012A&A...541A.153T,2013A&A...552A.102T}.
%
%Thereverse shocks have a large population
%of heavy elements by the supernova nucleosynthesis.
%The reverse shock acceleration \cite{2005JPhG...31R..95H}
%\cite{2005A&A...429..569E} has investigated a nonlinear
%particle acceleration at the reverse shock. The nonlinear effect can
%compress the shock \citep{1999ApJ...526..385B}.
%\citep{2012A&A...541A.153T,2013A&A...552A.102T}

Cassiopeia A with an age of $\sim350$ yr \citep{2006ApJ...645..283F}
is one of the most well-studied young Galactic SNRs.  Non-thermal
X-ray emission has been detected from both the outer rim and interior
positions
\citep[e.g.,][]{2000ApJ...528L.109H,2003ApJ...584..758V,2005ApJ...621..793B,2008ApJ...686.1094H,2009PASJ...61.1217M,2015ApJ...802...15G}.
In particular, the brightest non-thermal X-ray emission comes
from the western limb slightly interior
to the forward shock position and is also strongly associated with TeV
$\gamma$-ray emission
\citep[e.g.,][]{2001A&A...370..112A,2007A&A...474..937A,2009PASJ...61.1217M,2010ApJ...714..163A,2015ApJ...802...15G}.
Those features naturally suggest  efficient production of
accelerated particles in this region,
%
% There are also only a few samples of the SNRs which seem to
%accelerate particles in the reverse shock like Cassiopeia A: Kepler's
%SNR \citep{2002ApJ...580..914D}, RCW 86 \citep{2002ApJ...581.1116R}.
%In these samples, Cassiopeia A has the brightest non-thermal X-rays
%and .
however, the origin of the energetic electrons in the interior of the
remnant remains a mystery.

Radio and X-ray observations of Cassiopeia A have found that some
features are moving back toward the expansion center
\citep{1995ApJ...441..307A,1996ApJ...466..309K,2004ApJ...613..343D};
we refer to these as ``inward shocks''. The inward shocks are
concentrated in the south and west between azimuths of 170$^{\circ}$
to 300$^{\circ}$.  \cite{2008ApJ...686.1094H} have suggested that the
bright non-thermal X-rays at the reverse shock region could be due to
the locally higher reverse shock velocity($v_s \sim$ 6,000 km
s$^{-1}$) of the inward moving shock filaments.
%
% Also, \cite{2008ApJ...677L.105U} found the year-scale flux
%variabilities at the reverse shock regions, which implies an largely
%amplified magnetic field of $\sim1$ mG.
%
It was recognized even earlier that such high reverse shock velocities
or inward moving filaments required a more complex structure for the
circumstellar medium in the west.  For example,
\cite{1996ApJ...466..309K} argued for a physical connection with a
local molecular cloud based on strong radio and X-ray absorbing
structures seen near the inward moving shocks \citep[see also][for the
  X-ray absorbing
  structures]{2006MNRAS.371..829S,2012ApJ...746..130H}.  Recent
millimeter observations have additionally suggested an interaction
between the shock front and a nearby molecular cloud
\citep{2014ApJ...796..144K}. Such a high density cloud would also be
consistent with the pion-decay scenario for the $\gamma$-ray emission
from Cassiopeia A. Finally, a shock-cloud interaction is considered to
have a connection with the non-thermal X-rays
\citep{2010ApJ...724...59S,2013ApJ...778...59S,2015ApJ...799..175S,2012ApJ...744...71I}.

%On the other hand, \cite{2009ApJ...697..535P} argued...
%Another interesting feature at the western region is
%an existence of ``inward shock''.

Although these various non-thermal features appear to be related, a
unified picture for the western reverse shock region of Cassiopeia A
has yet to be established.  As a step in further understanding the
particle acceleration properties in this exceptional region of the
remnant, here we report new X-ray kinematic and intensity properties
for bright inward moving filaments.
%
%In order to understand how those are related
%physically, we investigate their properties in more detail.
% In this paper, we focus on the X-ray studies of the inward shocks.
%
Using long-term monitoring data from {\it Chandra}, we determine the
proper motion and temporal flux variation for a set of six bright
non-thermal filaments.  We then combine this with results from hard
X-ray measurements by {\it NuSTAR} to determine the physical
properties (e.g., magnetic field, diffusion coefficient, maximum
electron energy, and acceleration and cooling timescales) of the
particle acceleratio process in these shocks.

\section{Observation and Data Reduction} \label{sec:obs}
\subsection{\it Chandra}
The {\it Chandra} ACIS-S have observed Cassiopeia A several times since the launch in 1999
\citep*[e.g.,][]{2000ApJ...537L.119H,2004ApJ...615L.117H,2011ApJ...729L..28P,2012ApJ...746..130H,2014ApJ...789..138P}.
The data used in our analysis are listed in Table \ref{tab:chandradata}.
We reprocessed the event files (from level 1 to level 2) to remove pixel randomization
and to correct for CCD charge transfer efficiencies using CIAO \citep{2006SPIE.6270E..1VF} version 4.6 and CalDB 4.6.3.
The bad grades were filtered out and good time intervals were reserved.

\begin{table}[h!]
\scriptsize
\caption{{\it Chandra} observations.}
\begin{center}
\begin{tabular}{lcccc}
\hline
ObsID. 	 & 	date 		&	 exp. 			&		angle 			&		shift values								\\
		 &	YYYY/MM/DD 	&	(ks)			&		(deg)			&	($\Delta x$, $\Delta y$)\tablenotemark{1}		\\ \hline
114 	.& 	2000/01/30 	&	 49.9 			&		323.4			&		($-0.09$, $+0.34$)							\\ \hline
1952 	.& 	2002/02/06 	&	 49.6 			&		323.4			&		($-0.36$, $+0.38$)							\\ \hline
4634 	.& 	2004/04/28 	&	148.6 			&		 59.2			&		($+0.15$, $-0.20$)							\\
4635 	.& 	2004/05/01 	&	135.0 			&		 59.2			&		($-0.05$, $-0.27$)							\\
4636 	.& 	2004/04/20 	&	143.5 			&		 49.8			&		($-0.09$, $-0.21$)							\\
4637 	.& 	2004/04/22 	&	163.5 			&		 49.8			&		($-0.04$, $-0.22$)							\\
4638 	.& 	2004/04/14 	&	164.5 			&		 40.3			&		(reference)									\\
4639 	.& 	2004/04/25 	&	 79.0 			&		 49.8			&		($-0.04$, $-0.35$)							\\
5196 	.& 	2004/02/08 	&	 49.5 			&		325.5			&		($+0.07$, $+0.03$)							\\
5319 	.& 	2004/04/18 	&	 42.3 			&		 49.8			&		($-0.08$, $+0.08$)							\\
5320 	.& 	2004/05/05 	&	 54.4 			&		 65.1			&		($+0.03$, $-0.41$)							\\ \hline
9117 	.& 	2007/12/05 	&	 24.8 			&		278.1			&		($-0.01$, $+0.87$)							\\
9773 	.& 	2007/12/08 	&	 24.8 			&		278.1			&		($-0.19$, $+0.89$)							\\ \hline
10935 	.& 	2009/11/02 	&	 23.3 			&		239.7			&		($-0.53$, $+0.53$)							\\
12020 	.& 	2009/11/03 	&	 22.4 			&		239.7			&		($-0.50$, $+0.32$)							\\ \hline
10936 	.& 	2010/10/31 	&	 32.2 			&		236.5			&		($-0.44$, $-0.11$)							\\
13177 	.& 	2010/11/02 	&	 17.2 			&		236.5			&		($+0.00$, $-0.02$)							\\ \hline
14229 	.& 	2012/05/15 	&	 49.1 			&		 75.4			&		($-0.52$, $-0.60$)							\\ \hline
14480 	.& 	2013/05/20 	&	 48.8 			&		 75.1			&		($-0.38$, $+0.63$)							\\ \hline
14481 	.& 	2014/05/12 	&	 48.4 			&		 75.1			&		($+0.08$, $+0.50$)							\\ \hline
\end{tabular}
\label{tab:chandradata}
\end{center}
\tablenotetext{1}{Sky pixel value (1 pixel = 0.492$^{\prime\prime}$). }
\end{table}

In order to align the images, we used the Central Compact Object (CCO) as a fiducial source.
Because of lack of other certain fiducial sources around Cassiopeia A, we registered only the CCO for the alignment.
First, we determined the CCO position using \verb"wavdetect" implemented in CIAO\footnote{http://cxc.harvard.edu/ciao/}.
Next, shift values between two ObsIDs were computed with \verb"wcs_match" script. By using these values,
we finally updated the aspect solutions of the event files using \verb"wcs_update". We here used the ObsID 4638
as the reference of the alignment. The shift values are summarized in Table \ref{tab:chandradata}. In the following
section \ref{sec:pro}, we assumed official systematic errors of $\sim$0.5$^{\prime\prime}$\footnote{See http://cxc.harvard.edu/cal/ASPECT/celmon/ for the Chandra absolute astrometric accuracy.}
for the positional accuracy although we aligned the images. Our image alignment using the CCO could reduce the uncertainties. However, there would
still be other positional uncertainties (e.g., roll angle, proper motion of the CCO, etc.).
For example, \cite{2006ApJ...645..283F} have estimated a transverse velocity of $\simeq$350 km s$^{-1}$ for the CCO.
This corresponds to a shift of $\sim$0.3$^{\prime\prime}$ ($<$ 1 pixel) for 14 years at the distance of 3.4 kpc \citep[see][for the distance]{1995ApJ...440..706R},
which is within the systematic errors.

\begin{table}[h!]
\caption{{\it NuSTAR} observations.}
\begin{center}
\begin{tabular}{lcccc}
\hline
ObsID. 	 		& 	Date 		&	 Exp. 			\\
		 		&	YYYY/MM/DD 	&	(ks)			\\ \hline
40001019002 	& 	2012/08/18 	&	 294 			\\
40021001002 	& 	2012/08/27 	&	 190 			\\
40021001004 	& 	2012/10/07 	&	 29 			\\
40021001005		& 	2012/10/07 	&	 228 			\\
40021002002 	& 	2012/11/27 	&	 288 			\\
40021002006 	& 	2013/03/02 	&	 160 			\\
40021002008 	& 	2013/03/05 	&	 226 			\\
40021002010 	& 	2013/03/09 	&	 16 			\\
40021003002 	& 	2013/05/28 	&	 13 			\\
40021003003 	& 	2013/05/28 	&	 216 			\\
40021011002 	& 	2013/10/30 	&	 246 			\\
40021012002 	& 	2013/11/27 	&	 239 			\\
40021015002 	& 	2013/12/21 	&	 86 			\\
40021015003 	& 	2013/11/23 	&	 160 			\\ \hline
Total 			& 			 	&	 $\sim$2.4 Ms  	\\ \hline
\end{tabular}
\label{tab:nudata}
\end{center}
%\tablenotetext{1}{Sky pixel value (1 pixel = 0.492$^{\prime\prime}$). }
\end{table}
\subsection{\it NuSTAR}

{\it NuSTAR}, which is composed of two co-aligned X-ray telescopes (FPMA and FPMB) observing
the sky in the energy range 3--79 keV \citep*[][]{2013ApJ...770..103H}, has observed Cassiopeia A
with a total of $\sim$ 2.4 Ms of exposure time during the first two years after the launch in 2012
(see Table \ref{tab:nudata}). The first hard X-ray imaging revealed $^{44}$Ti distribution and
morphology of $>$15 keV emission in Cassiopeia A for the first time \citep*[][]{2014Natur.506..339G,2015ApJ...802...15G,2017ApJ...834...19G}.

\begin{figure*}[t]
 \begin{center}
  \includegraphics[bb=0 0 1150 1614,width=14cm]{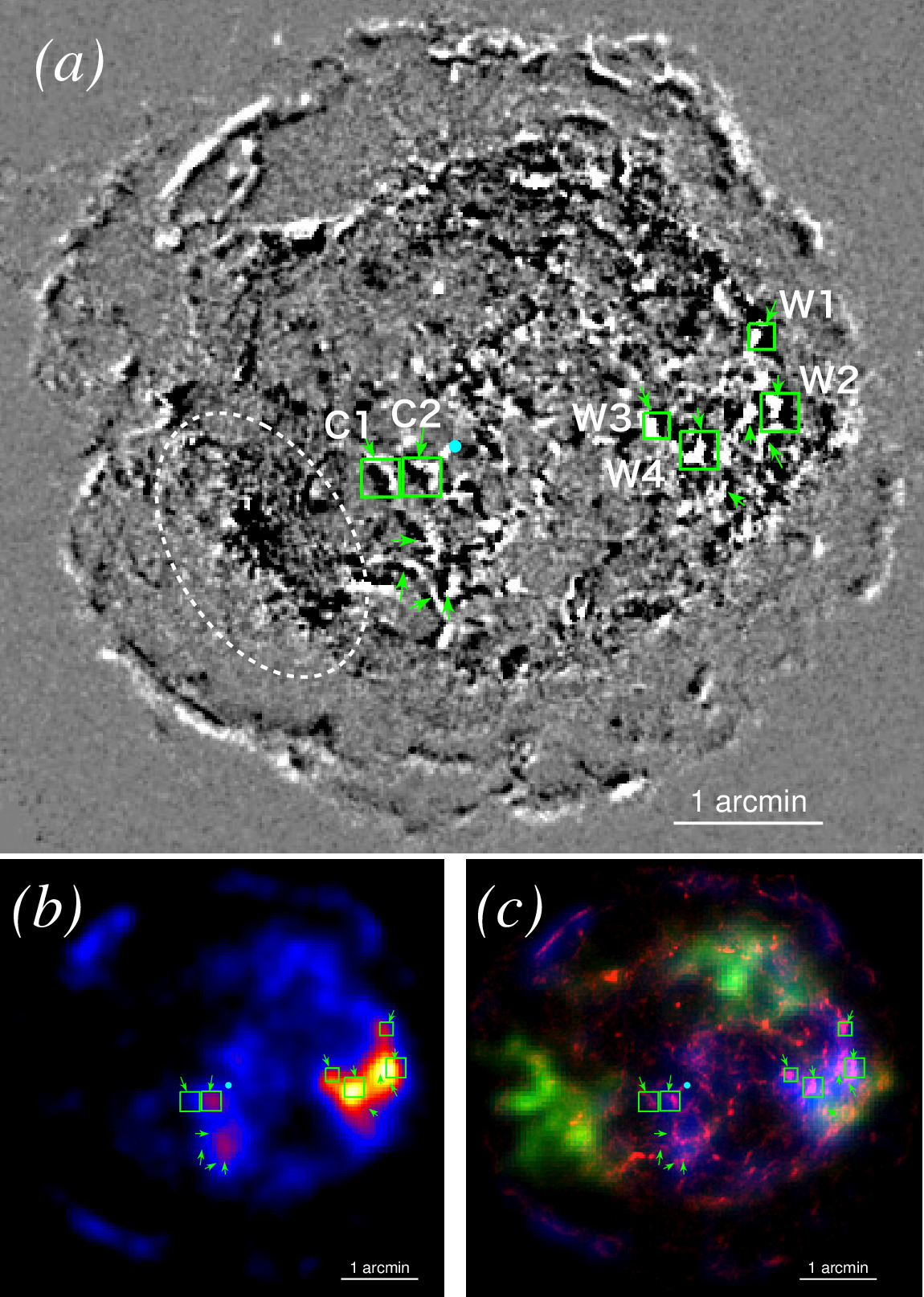}
 \end{center}
\caption{(a) Image difference in the 4.2--6 keV band of Cassiopeia A between 2000 and 2014 with {\it Chandra}.
Adjacent black and white features show transverse motions of X-ray structures. Green arrows and boxes show
the inward-shock positions defined by eye and the regions which were used for the proper-motion measurements.
The boxes are identified by name (e.g., C1, W1)
The box sizes are 21$\times$21 pixels (1 pixel = 0.492$^{\prime\prime}$) for W1 and W3, and 31$\times$31 pixels
for C1, C2, W2 and W4, respectively.
(b) The {\it NuSTAR} image in 15--40 keV band. The green arrows and boxes show the same as shown in the figure (a).
(c) Three-color image of Cassiopeia A. Red, green and blue color show the 4.2--6 keV image with {\it Chandra},
the 6.54--6.92 keV image (around Fe-K emission) with {\it Chandra} and the 15--40 keV image with {\it NuSTAR}.
The binning sizes for the 4.2--6 keV ({\it Chandra}), 6.54--6.92 keV ({\it Chandra}) and 15--40 keV ({\it NuSTAR})
images are 1$\times$1 pixels (0.492$^{\prime\prime}$$\times$0.492$^{\prime\prime}$), 8$\times$8 pixels
(3.936$^{\prime\prime}$$\times$3.936$^{\prime\prime}$) and 1$\times$1 pixels (12.3$^{\prime\prime}$$\times$12.3$^{\prime\prime}$),
respectively. The images were then smoothed with a Gaussian function with a sigma of 3 bins. The green arrows and boxes show
the same as shown in the figure (a). Cyan dot shows the position of the central compact object (CCO).}
\label{fig:diff}
\end{figure*}

\begin{figure*}[t]
 \begin{center}
  \includegraphics[bb=0 0 1938 1614,width=17cm]{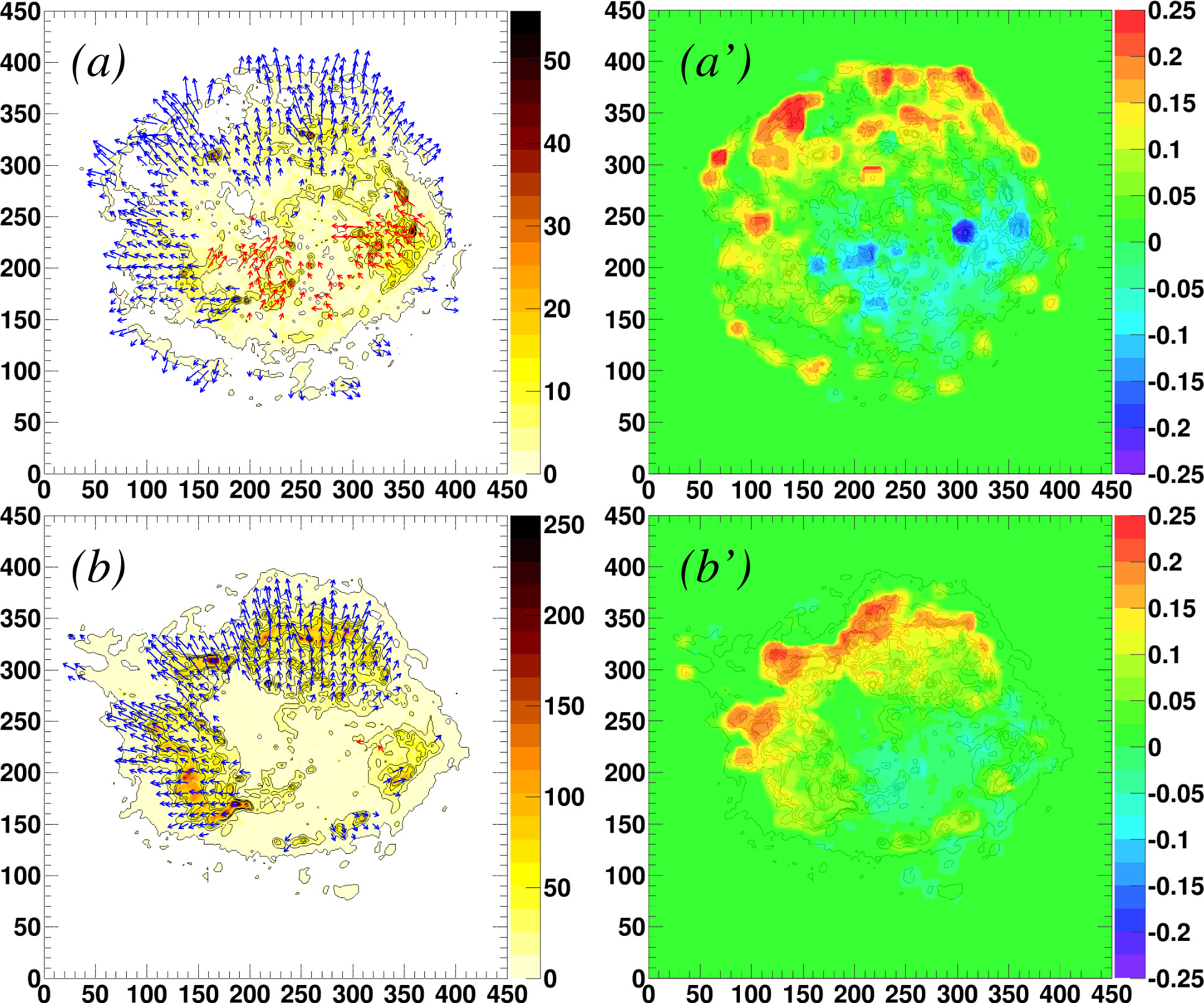}
 \end{center}
\caption{{\it Left side}: X-ray images overlaid with proper-motion vectors obtained by the optical flow method, where the energy
ranges are (a) a continuum band of 4.1-6.3 keV and (b) Si-K band of 1.7-2.1 keV. For the optical flow analysis,
we used images taken in 2004 (ObsID. 5320) and 2014 (ObsID. 14481). The vector length is proportional to the actual shifting value.
Blue and red show outward and inward motions, respectively. In these figures, small vectors
whose proper motion is $\lesssim$ 0.05 arcsec yr$^{-1}$ are not shown. The field of view is 450$\times$450 pixels
(= 3.69$^\prime$$\times$3.69$^\prime$). The center of the field of view, (x, y) = (225, 225) is the CCO location.
The unit of the color bar is 10$^{-7}$ counts cm$^{-2}$ s$^{-1}$. Contours overlaid on the continuum
and Si-K band images show 1, 5, 10, 15, 20$\times$10$^{-7}$ counts cm$^{-2}$ s$^{-1}$ and 3, 15, 30, 45, 60$\times$10$^{-7}$ counts
cm$^{-2}$ s$^{-1}$, respectively.
{\it Right side}: length of radial component of proper-motion vectors are shown in color maps
in the continuum band (a$^{\prime}$) and the Si-K band (b$^{\prime}$). Positive and negative values correspond to outward and inward motions, respectively.
The unit of the color scale is arcsec yr$^{-1}$. A proper motion of 0.15 arcsec yr$^{-1}$ corresponds to $\sim$2,400 km s$^{-1}$ at
the distance of 3.4 kpc.}
\label{fig:diff2}
\end{figure*}

In order to compare the hard X-ray image (15--40 keV) with the {\it Chandra} image (4.2--6 keV), we analyzed all of the {\it NuSTAR} data.
We reduced the data with the {\it NuSTAR} Data Analysis Software (NuSTARDAS) version 1.6.0 and {\it NuSTAR} calibration database
(CALDB) version 20170503 to produce images and exposure maps for each telescope. All of the images were
merged by using \verb"XIMAGE", taking into account the exposure maps (i.e., the unique time-dependent
exposure and the vignetting function of the telescope) for each pointing. After the correction, we deconvoluved
the merged image with the on-axis PSF image of the {\it NuSTAR}'s telescope. We used \verb"arestore" script in
CIAO for the deconvolution. Following the method in \cite{2015ApJ...802...15G}, we chose to halt the deconvolution
after 50 iterations with a Richardson-Lucy method, and we obtained almost the same image as the published image
(see Figure \ref{fig:diff}). For the {\it NuSTAR} imaging, the default aspect solution was used.

\section{Analysis and Results} \label{sec:result}

\subsection{Soft and Hard X-ray imaging} \label{sec:pro}

In Figure \ref{fig:diff}(a), we show a difference image between 2000 and 2014 generated from {\it Chandra} data.
Here we can see moving X-ray structures as previously reported \citep[e.g., ][]{2004ApJ...613..343D,2009ApJ...697..535P}.
All of the forward-shock filaments which are located at the outer rim are moving outward. On the other
hand, there are several filaments near the center and toward the west that are moving inward (such filaments are identified
with green arrows in the figure). Those filaments are the same inward-moving features (inward shocks) as reported in \cite{2004ApJ...613..343D}.
In particular, almost all the filaments toward the west seem to be inward shocks. By definition, the forward shock must
always move to the outside of the remnant. Therefore, these inward shocks cannot be forward shocks seen projected on
the interior of the remnant. In addition, part of the eastern region shows a uniform dark color (see a
broken ellipse region in Figure \ref{fig:diff}). The eastern region is known to have a large contribution from thermal
X-rays with a flux decline possibly due to adiabatic expansion \citep{2017ApJ...836..225S}. The dark eastern region in this
figure is just another view of the prominent flux decay of the thermal emission.

In Figure \ref{fig:diff}(b), we show the 15--40 keV image with {\it NuSTAR}. As reported in previous studies
\citep{2002A&A...381.1039W,2008ApJ...686.1094H,2009PASJ...61.1217M,2015ApJ...802...15G,2017ApJ...836..225S},
most of the hard X-rays are located at the west and the center. Thus, we find a strong correlation
between the inward-shock positions and the intensity peaks of the {\it NuSTAR} image.

In Figure \ref{fig:diff}(c), we compared the {\it NuSTAR} image with the soft X-ray images (4.2--6 keV: continuum emission \& 6.54--6.92
keV: Fe-K emission) with {\it Chandra}. As reported in \cite{2017ApJ...836..225S}, we can see a separation between the thermal X-ray
distribution (Fe-K) and the hard X-ray distribution (which is spatially coincident with the inward-shock positions).
%\software{NuSTARDAS (v1.6.0), CIAO (Fruscione et al. 2006), OpenCV (v3.2.0;  [http://opencv.org] /), XSPEC (v12.9.0n; Arnaud 1996)}

To clarify the motions of the shock filaments, we also compared images taken in 2004 (ObsID. 5320) and 2014 (ObsID. 14481) using a
computer vision technique (optical flow). The dense optical flow algorithm we use is by  Gunnar Farneb{\"a}ck \citep{Farneback_2003} as
implemented in OpenCV 3.2.0\footnote{See http://opencv.org/ for more details.}. This algorithm estimates displacement fields between
two frames using quadratic polynomials to locally approximate the image flux. To reduce the noise, the estimate is made by averaging over local
neighborhoods. Also, we here set statistical limits for the surface brightnesses. For the continuum and Si-K images,
we ignore pixels whose surface brightness is $<$ 1$\times$10$^{-7}$ counts cm$^{-2}$ s$^{-1}$ and $<$ 3$\times$10$^{-7}$ counts cm$^{-2}$ s$^{-1}$
respectively, which corresponds to $\lesssim$ 1 counts with $\sim$ 50 ksec exposure time. Specifically, we used the \verb"calcOpticalFlowFarneback"
function with the following arguments: \verb"pyr_scale" = 0.5,
which specifies the image scale to build pyramids for each image, \verb"levels" = 3, number of pyramid layers, \verb"winsize" = 15,
averaging window size, and \verb"poly_n" = 5, size of the pixel neighborhood used for the polynomial approximation.
As shown in Figure \ref{fig:diff2}, we can visualize the motion of filaments (or small knots) as vector maps over the entire extent of the SNR.
In the vector maps (Figure \ref{fig:diff2} (a) \& (b)), we plot only large vectors whose proper motion is larger than
$\sim$ 0.05 arcsec yr$^{-1}$ ($\approx$ 800 km s$^{-1}$ at the distance of 3.4 kpc), and we here found the large motions are strongly associated
with filamentary (or knotty) structures. Also, we can see clearly that large motions are not identified at low count-rate regions.

In the continuum band (Figure \ref{fig:diff2} (a) \& (a')), the inward motions are concentrated in the interior. Indeed the positions of
inward-moving features (blue spots in \ref{fig:diff2} (a')) are very similar to the locations of the hard X-ray spots in Figure \ref{fig:diff}(b),
providing further  support for a connection between them. On the other hand, almost all of the small structures in the Si-K band are moving to the
outside of the remnant (Figure \ref{fig:diff2} (b) \& (b')). These indicate a clear difference in the character of motion between the ejecta and the
non-thermal emission components.

In this current study, we employ the optical flow method to provide a
largely qualitative view of the expansion of the thermal ejecta and
non-thermal shock filaments in Cassiopeia A.  A detailed quantitative
study of the results is beyond the scope of this initial
investigation, although in future work, we will determine the accuracy
and robustness of optical-flow proper-motion measurements for
determining the global kinematics of young SNRs.  In particular,
combining comprehensive proper-motion and radial velocity measurements
will further improve our understanding of the three-dimensional kinetics of
young SNRs \citep[e.g.,][]{2010ApJ...725.2038D,2017ApJ...840..112S,2017ApJ...842...28W,2017ApJ...845..167S}.

%To clarify the motions of the shock filaments, we also compared images taken in 2004 (ObsID. 4638) and 2014 (ObsID. 14481) by using a
%computer vision technique, optical flow. We applied a dense optical flow using the Gunnar Farneback's algorithm \cite{Farneback_2003},
%implemented in OpenCV 3.2.0\footnote{See http://opencv.org/ for more details.}.

\begin{table*}[t!]
\scriptsize
\caption{Proper Motions and Spectral Parameters of the Inward-Shock Filaments in Cassiopeia A\tablenotemark{1}}\label{tab:proper}
\begin{center}
\begin{tabular}{lccccc|cc}
\hline
		&																							\multicolumn{5}{c|}{\bf Imaging Analysis}																					&	\multicolumn{2}{c}{\bf Spectral Parameters in 2004}				\\
		&		center of the model frame 															&		mean shift: $\Delta$x, $\Delta$y		&	proper motion\tablenotemark{2}		&		velocity\tablenotemark{3}					&	angle 			&	Photon Index				&	flux in 4.2-6 keV	\\
id		&		 R.A., Decl. (epoch 2004)															&			(arcsec yr$^{-1}$)					&	(arcsec yr$^{-1}$)	&		(km s$^{-1}$)								&	(degree) 		&								&	(10$^{-13}$ ergs cm$^{-2}$ s$^{-1}$)	\\ \hline
\multicolumn{2}{l}{\bf west region}																	&												&						&													&					&								&						\\
W1		& 23$^{\rm h}$23$^{\rm m}$12$^{\rm s}$.070, $+58^\circ$49$^\prime$23$^{\prime\prime}$.64	&	$-0.111\pm$0.026, $+0.077\pm$0.027			&	0.132$\pm$0.027		&			2130$\pm$440							&	145$\pm$11 		&	2.11$^{+0.05}_{-0.08}$		&	5.70$\pm$0.04		\\
%		& 																							&	  ($\sigma_{\rm SD}$ = 0.004, 0.010)		&						&													&					&								&						\\
W2		& 23$^{\rm h}$23$^{\rm m}$11$^{\rm s}$.126, $+58^\circ$48$^\prime$56$^{\prime\prime}$.07	&	$-0.180\pm$0.026, $-0.089\pm$0.027			&	0.200$\pm$0.026		&			3210$\pm$420							&	206$\pm$8 		&	2.00$^{+0.03}_{-0.05}$		&	19.79$\pm$0.07		\\
%		& 																							&	($\sigma_{\rm SD}$ = 0.004, 0.010)			&						&													&					&								&						\\
W3		& 23$^{\rm h}$23$^{\rm m}$17$^{\rm s}$.525, $+58^\circ$48$^\prime$50$^{\prime\prime}$.73	&	$-0.190\pm$0.027, $-0.113\pm$0.027			&	0.219$\pm$0.027		&			3540$\pm$440							&	211$\pm$7 		&	2.18$^{+0.05}_{-0.08}$		&	6.65$^{+0.05}_{-0.03}$		\\
%		& 																							&	($\sigma_{\rm SD}$ = 0.004, 0.010)			&						&													&					&								&						\\
W4		& 23$^{\rm h}$23$^{\rm m}$15$^{\rm s}$.310, $+58^\circ$48$^\prime$40$^{\prime\prime}$.87	&	$-0.126\pm$0.026, $-0.042\pm$0.027			&	0.129$\pm$0.026		&			2090$\pm$420							&	199$\pm$12 		&	1.94$^{+0.04}_{-0.07}$		&	10.04$^{+0.06}_{-0.08}$		\\
%		& 																							&	($\sigma_{\rm SD}$ = 0.004, 0.010)			&						&													&					&								&						\\
\multicolumn{2}{l}{\bf center region}																&												&						&													&					&								&						\\
C1		& 23$^{\rm h}$23$^{\rm m}$31$^{\rm s}$.829, $+58^\circ$48$^\prime$29$^{\prime\prime}$.70	&	$+0.215\pm$0.026, $-0.013\pm$0.027			&	0.213$\pm$0.027		&			3440$\pm$440							&	356$\pm$7 		&	2.64$^{+0.06}_{-0.09}$		&	6.56$^{+0.08}_{-0.05}$		\\
%		& 																							&	($\sigma_{\rm SD}$ = 0.004, 0.010)			&						&													&					&								&						\\
C2		& 23$^{\rm h}$23$^{\rm m}$29$^{\rm s}$.676, $+58^\circ$48$^\prime$30$^{\prime\prime}$.20	&	$+0.214\pm$0.027, $+0.101\pm$0.027			&	0.235$\pm$0.027		&			3800$\pm$440							&	25$\pm$6 		&	2.24$^{+0.05}_{-0.07}$		&	8.41$^{+0.05}_{-0.07}$		\\
%		& 																							&	($\sigma_{\rm SD}$ = 0.004, 0.010)			&						&													&					&								&						\\
\hline
\end{tabular}
\end{center}
\tablenotetext{1}{All of the errors in this table show 68 \% confidence level ($\Delta \chi^2$ = 1.0). For the image analyses, the systematic uncertainties are also included.}
\tablenotetext{2}{The proper motions and the errors are estimated by using the Rice distribution (see Appendix).}
\tablenotetext{3}{The distance to the remnant was assumed to be 3.4 kpc.}
\end{table*}

\subsection{Proper-Motion Measurements of the Inward-Shock Filaments} \label{sec:pro}

\begin{figure*}[h!]
 \begin{center}
  \includegraphics[bb=0 0 2095 1560,width=17cm]{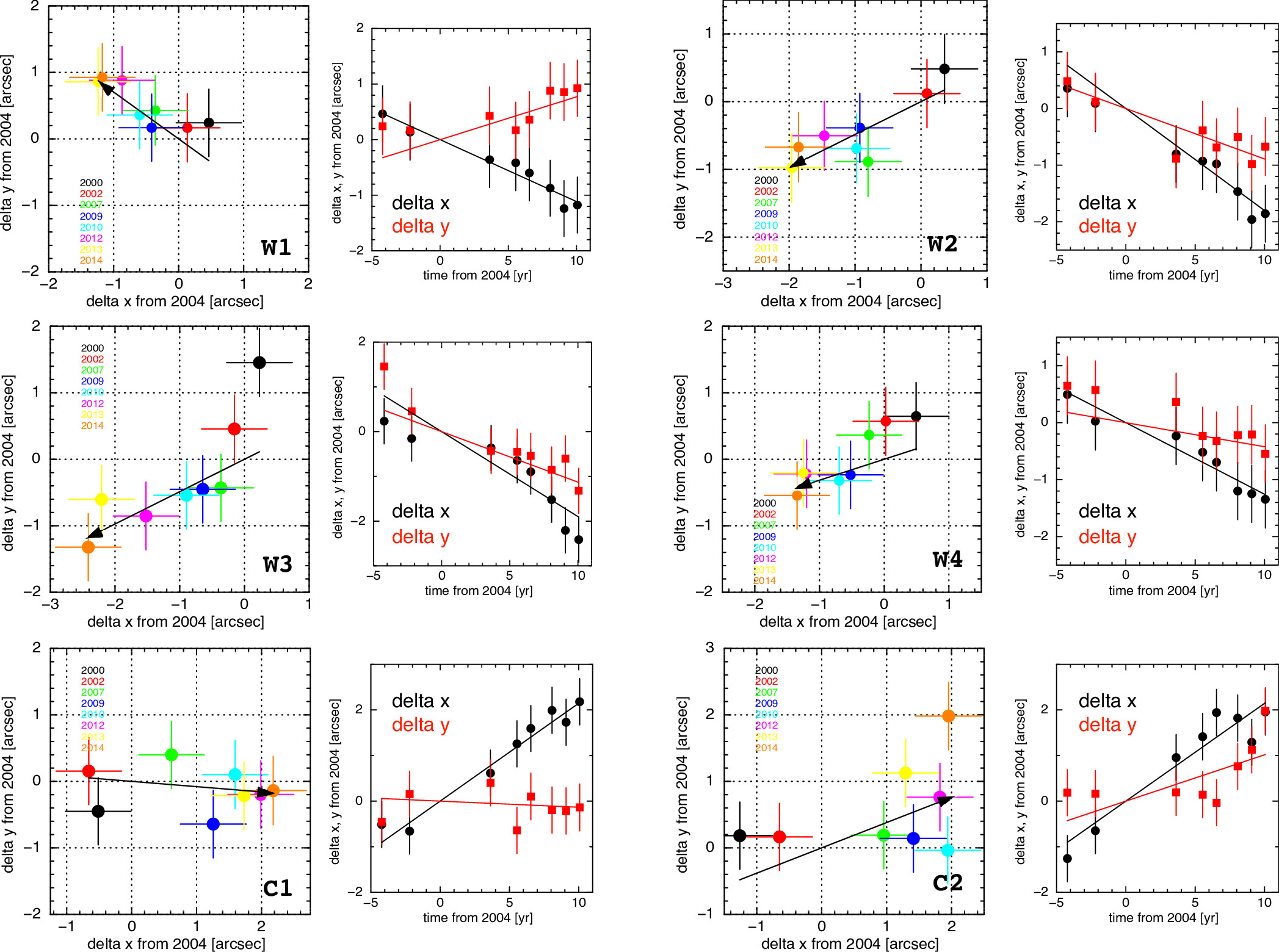}
 \end{center}
\caption{Results of the proper-motion measurements. Large panel shows a scatter plot of the best-fit positions (delta x vs. delta y).
Here, each color identifies the best-fit positions from 2004. Black arrows show best-fit directions of the proper motions. Small panel that is placed
close to the large one shows a plot of the best-fit positions as a function of time (delta x, y vs. time). Solid lines show the best-fit linear models
for the estimations of the mean shift values summarized in Table \ref{tab:proper}.
The error bars show 68 \% confidence level, which include the systematic errors of
($\sigma_{\rm x}$, $\sigma_{\rm y}$) $=$ (0.5$^{\prime\prime}$, 0.5$^{\prime\prime}$). The fitting (statistical) errors are much smaller than the
systematic uncertainties.}
\label{fig:pro_fit}
\end{figure*}

To measure proper motions of the inward-shock filaments, we extracted images from the box regions in
Figure \ref{fig:diff} and fitted with an image model. The fitting code used for this analysis was originally
developed for proper-motion measurements for knots in Kepler's SNR \citep{2017ApJ...845..167S}.
Here, we used the 2004 image normalized by the image in each epoch as the model because the observation
in 2004 has the longest exposure time that is about 20 times longer than that of the other observations (see Table \ref{tab:chandradata}).
For obtaining the best-fit shifts, a
maximum likelihood statistic for Poisson distributions was used \citep[C-statistics; ][]{1979ApJ...228..939C},
which minimizes
\begin{equation}
 C = -2{\rm ln}~P = -2\Sigma_{i,j}(n_{i.j}~{\rm ln}~m_{i,j}-m_{i,j}-~{\rm ln}~n_{i.j}!)
\end{equation}
where $n_{i,j}$ is the counts in pixel ({\it i,j}) of the image in each epoch, and $m_{i,j}$ is
the model counts based on the 2004 image. The fitting errors can be estimated in the usual manner since the statistical distribution of
 $\Delta C = C - C_{\rm min}$ is similar to that of $\chi^2$ \citep{1979ApJ...228..939C}.

The fitting results are summarized in Figure \ref{fig:pro_fit} and Table \ref{tab:proper}.
%As a result, we got reasonable fitting results: %reduced $C =$ 1.51--2.07 for W1, 1.63--2.56 for W2,
%1.32--2.04 for W3, 1.28--2.05 for W4, 1.20--1.41 for C1, 1.38--1.62 for C2.
We found that the best-fit
positions of those filaments are gradually shifting from the outside to the inside with large proper motion
values: 0.129--0.219$^{\prime\prime}$ yr$^{-1}$ for the W filaments and
0.213--0.235$^{\prime\prime}$ yr$^{-1}$ for the C filaments.
The best-fit x, y shifts and errors listed in Table \ref{tab:proper} were determined
by chi-square fitting with a linear function as shown in Figure \ref{fig:pro_fit}. In the fitting,
the slope provides the x, y shift (arcsec per year) with the error coming from $\Delta \chi^2$ = 1.0.
Using the values, we determined the proper motions and errors (see Appendix for the method).
% Here, we estimated the unbiased proper motions and errors
%using the Rice distribution (see Appendix).}
For the assumed distance of 3.4 kpc
\citep{1995ApJ...440..706R} to the remnant, we  estimate the filament velocities to be $\sim$2,100--3,800
km s$^{-1}$.

In the radio band, the kinematics of 304 knot structures over the
entire remnant had been reported previously
\citep{1995ApJ...441..307A}.  This work showed that the motions of
many western knots deviated from that of knots in other regions and
that some western knots were actually moving inward \citep[see Fig. 5
  and Fig. 6 in][]{1995ApJ...441..307A}.  In the X-ray band,
continuum-dominated filaments with an inward motion were also found at
azimuths between 170$^{\circ}$ and 300$^{\circ}$
\citep{2004ApJ...613..343D}. These features correspond to the inward
shocks we found. %\textcolor{red}{\bf Coming soon...}

\subsection{Flux Variations} \label{sec:flux}
\begin{figure*}[t]
 \begin{center}
  \includegraphics[bb=0 0 2000 478,width=17cm]{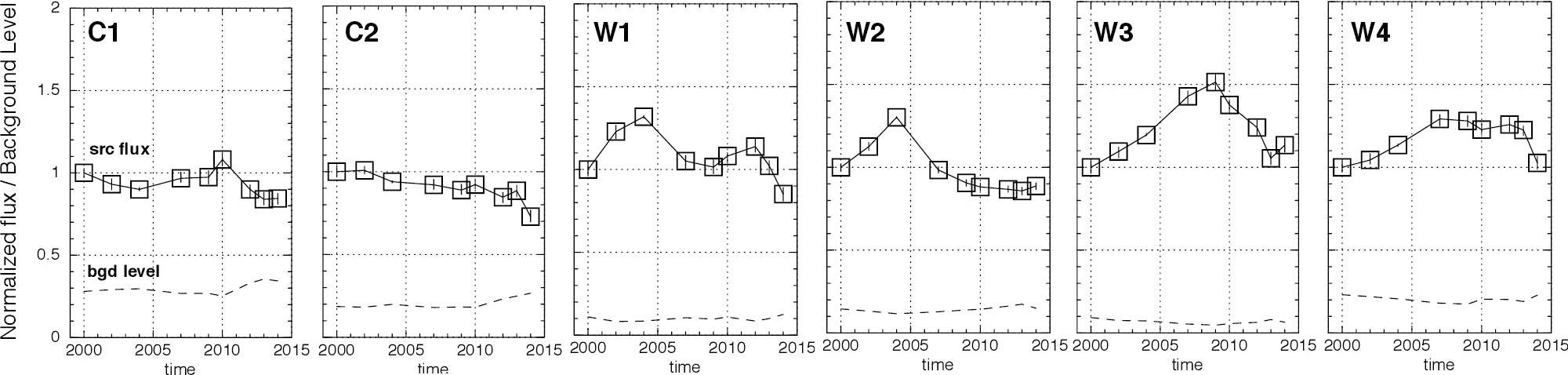}
 \end{center}
\caption{Time variations of the flux (black boxes) for each filament in the 4.2-6 keV band.
The fluxes were normalized by the flux in the first epoch (2000). The error bars show 68 \% confidence level ($\Delta \chi^2$ = 1.0).
Broken lines show background levels in each epoch.}
\label{fig:flux}
\end{figure*}

Here we investigate the X-ray light curves of the inward moving filaments. First we extracted spectra from each epoch for each inward-shock filament
using the CIAO \verb"specextract" command. For the source regions, we used ellipse regions which included the bright filament
structures. These regions were shifted with time for taking the proper-motion effects into account. The background
regions were selected from nearby low-brightness regions. The background-subtracted spectra were fitted in the 4.2--6 keV
band using a \verb"power-law" model in Xspec 12.9.0n \citep{1996ASPC..101...17A}.
%A background level in the extracted spectrum is $\lesssim$ 30 \%,
%which has a fluctuation of $\lesssim \pm 20$ \% in 14 years (see black-broken lines in Figure \ref{fig:flux}).
%\textcolor{red}{The background fluctuation could change estimations of X-ray fluxes by $\sim$9\%.  Even if we take
%this effect into account, we found significant time variations of the X-ray flux.}
%Although the fluctuation could change an estimation of X-ray fluxes by $\sim$9 \%, we found significant time variations
%even taking the effects into account.

We summarize the fitting results in Figure \ref{fig:flux}. For all of the western filaments, over the $\sim$14 yr span of observation
X-ray fluxes increased by up to $\sim$40--50 \% and afterward decreased to the initial flux level from epoch 2000. Background photons in
the extracted spectrum are less than 30 \% of the total, and fluctuate by $\lesssim \pm 20$ \% (see black-broken lines in Figure \ref{fig:flux}).
The change in the background rate from a uniform level could introduce a fluctuation in the filament's X-ray flux of at most $\sim$9 \%.
Even given this effect, significant time variations of the X-ray flux are still present.
Time intervals of the monitoring observations would mainly influence on the estimation accuracy of variable timescales.
The intervals are $\sim$ 2 yrs from 2000 to 2007 and
$\sim$ 1 yr from 2007 to 2014. Therefore, the uncertainty of estimated timescales would be $\pm$ 2 yr for the W1 and W2 filaments
and $\pm$ 1 yr for the W3 and W4 filaments. Even taking these errors into consideration, the timescales are very short.
For regions W1 and W2, the timescales of the flux increase up to the maximum level ($t_{\rm in}$) and decrease
down to the initial level ($t_{\rm dec}$) are $t_{\rm in} \sim t_{\rm dec} = 4 \pm 2$ yr.
For regions W3 and W4, the increase times are longer than those at the W1 and W2 filaments. These filaments have
a similar light curve although the W4 filament's one shows a little complex shape. As a representative of these, the W3 filament shows
$t_{\rm in} = 9 \pm 1$ yr and $t_{\rm dec} = 4 \pm 1$ yr.
On the other hand, regions C1 and C2 show little to no sudden change of flux; rather their fluxes seem to be gradually decreasing
by $\sim$20--30 \% over $\sim$14 yr time span. The variability of the non-thermal X-ray flux is strongly related to the magnetic
field \citep[e.g.,][]{2007Natur.449..576U}. The differences of the variable timescales seem
be related to differences in the magnetic field strength at each region (this is discussed in section \ref{sec:dis2}).

\cite{2007AJ....133..147P} and \cite{2008ApJ...677L.105U} have already shown X-ray variations in small structures of the remnant using
the first three epochs of {\it Chandra} observations. Regions W1 and W2 are nearly the same as regions R4 in \cite{2007AJ....133..147P} and H in \cite{2008ApJ...677L.105U},
respectively. Therefore, these are follow-up results for the past observations, and then the time variations are almost consistent
with each other in the same time interval.

Spectral parameters in 2004 are summarized in Table \ref{tab:proper}. %We found that all of the inward shocks show a harder
%spectral index $\Gamma \sim$ 2.0--2.6 than the hard-X-ray (15--40 keV) spectral index $\Gamma \sim$ 3.35 in the central knots
%\citep{2015ApJ...802...15G}.
We were not able to detect significant time variation in the photon indices.

\begin{figure}[h!]
 \begin{center}
  \includegraphics[bb=0 0 1400 864,width=8cm]{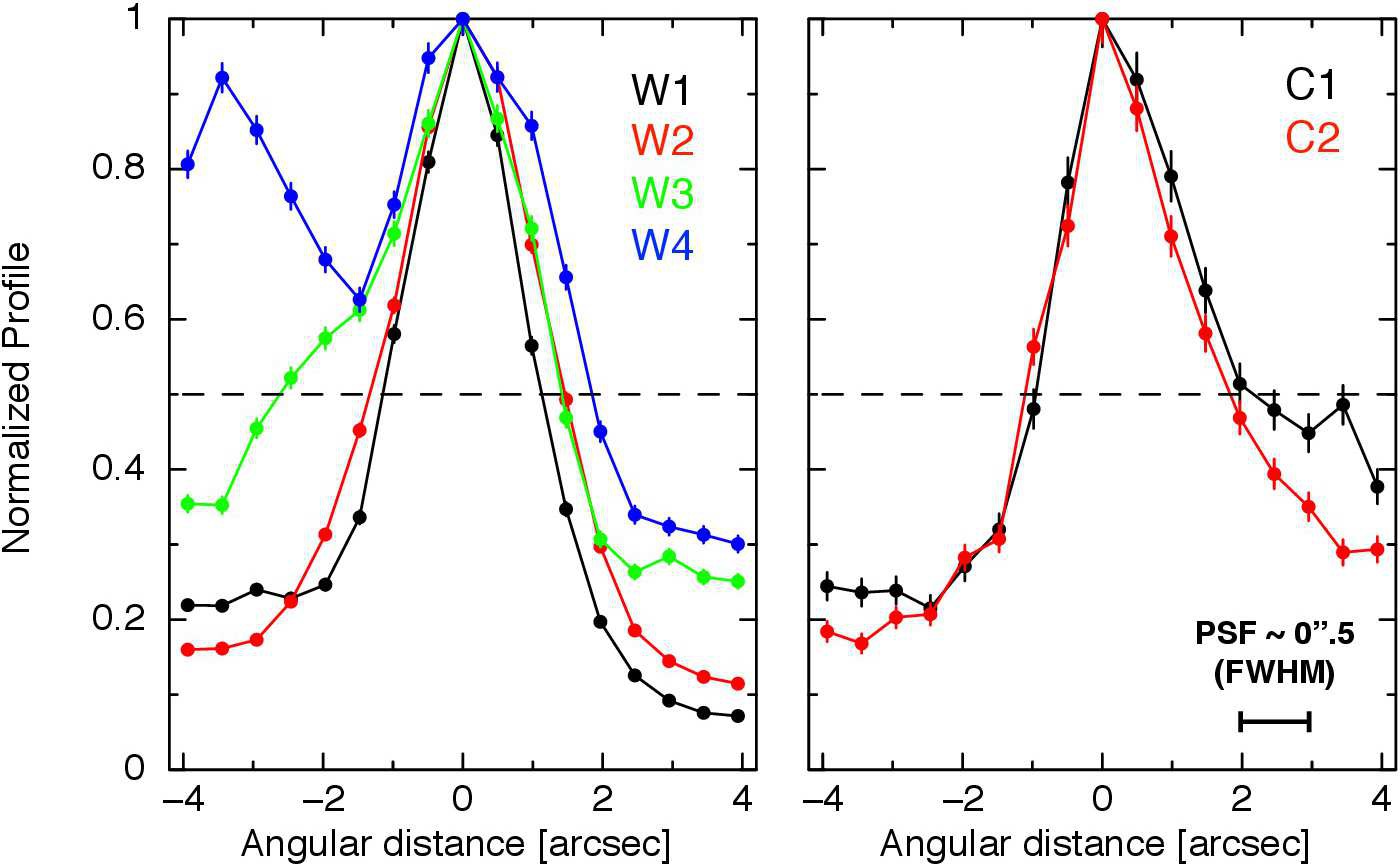}
 \end{center}
\caption{Emission profiles of the inward-shock filaments in the west (left) and center (right). The profiles were extracted from
15 pixels wide using 2004 image in 4.2--6 keV band and were normalized by the peak intensities. The positive direction in the angular
distance shows a direction to the outside of the remnant. The scale bar in the right panel shows the {\it Chandra}'s angular resolution,
$\sim$0$^{\prime\prime}$.5 (FWHM).}
\label{fig:width1}
\end{figure}

\subsection{Thickness of the Inward-shock Filaments} \label{sec:width}
We evaluate the spatial extent of the inward-shock filaments using {\it Chandra}'s high angular resolution of $\sim$0$^{\prime\prime}$.5 (FWHM).
We found all filament widths to be in the range 2$^{\prime\prime}$--4$^{\prime\prime}$ (Figure \ref{fig:width1}), which is very similar to
the widths of the forward-shock filaments \citep[1$^{\prime\prime}$.5--4$^{\prime\prime}$:][]{2003ApJ...584..758V}.
The angular resolution of the Wolter type-I optics used for the {\it Chandra}'s mirrors degrades at off-axis field angles. However, at the off-axis location of the inward-shock filaments (2$^{\prime}$--3$^{\prime}$), the angular resolution is nearly consistent
with that at the on-axis position\footnote{See http://cxc.harvard.edu/proposer/POG/html/chap4.html for the angular resolutions at off-axis angles.}.
Therefore, we are justified in ignoring the additional blur introduced by the off-axis effects for the shock thickness evaluation. At the remnant's distance of 3.4 kpc, the observational thickness is estimated to be $\sim$0.03--0.07 pc.
We can estimate a synchrotron loss time from the widths by combining them with the plasma fow speed away from the shock front (see section \ref{sec:dis2}
for this discussion).

\section{Discussion} \label{sec:discussion}
We have shown that some bright filaments are moving inward with velocities of $\sim$2,100--3,800 km s$^{-1}$ using all the monitoring data taken
by {\it Chandra} in 2000--2014. These filaments cannot be located at the forward-shock position even when projection is considered
because the forward shock must move outward. In addition, we found that those filaments are cospatial with  hard
X-ray hot spots seen by {\it NuSTAR} and additionally that these filaments show flux variations (both increasing and decreasing) on
timescales of a few years.
These results imply that the hard X-ray emission is related to non-thermal emission from accelerated electrons
at the inward-shock regions where the magnetic field is high amplified. These results hold the hope of advancing our understanding of the
mysterious non-thermal X-ray emission at the interior of Cassiopeia A and the particle acceleration that is taking place at these shocks.
In this section,
we discuss the implications of our results on the physical conditions at the inward shocks (section \ref{sec:dis1}), the process of
particle acceleration there
(section \ref{sec:dis2}) and the origin of the inward shocks themselves (section \ref{sec:dis3}).

\subsection{Inward Shock Conditions} \label{sec:dis1}
Here we make the plausible assumption based on the location of the inward shocks
(see Figure \ref{fig:diff}) that they are propagating through the inner ejecta.
Thus their actual shock velocities in the rest frame of the ejecta should be higher than what we  measure from the
proper motions. Here, we define the shock velocity $v_s$ as
\begin{equation}
 v_s \equiv v_{\rm ej} - v = \frac{R_{\rm ej}}{t} - v,
\end{equation}
where $v_{\rm ej}$, $v$, $R_{\rm ej}$ and $t$ are the ejecta velocity, the velocity we measure, the typical ejecta radius and the
age of the remnant.

First, we consider the case where the inward shocks are propagating into cool freely-expanding ejecta.
We adopt a value of $R_{\rm ej} \sim 5 \times 10^{18}$ cm, which was estimated from the typical reverse shock radius at the distance of 3.4 kpc
\citep{2001ApJ...552L..39G}, and then obtained an ejecta speed of $v_{\rm ej} \sim 4,900$ km s$^{-1}$ for $t \sim 320$ yr (the remnant's age in 2000).
The inward shocks have measured proper motion velocities $v$ from $-2,100$ km s$^{-1}$ to $-3,800$ km s$^{-1}$, leading to shock velocity estimates
of $v_{s} \sim 7,000$--$8,700$ km s$^{-1}$.
Next, we consider a case where the inward shocks are propagating into previously shocked, and therefore hot,  ejecta. In this case, $v_{\rm ej} $
corresponds to the bulk expansion velocity of the ejecta. \cite{2004ApJ...613..343D} estimated an expansion rate of 0.2 \% yr$^{-1}$ for the
ejecta component. Using this value and the same value of $R_{\rm ej}$ as above, the expansion velocity is estimated to be $\sim$3,000 km s$^{-1}$
\citep[see also][for the ejecta velocity]{1994PASJ...46L.151H,2002A&A...381.1039W,2004ApJ...614..727M,2010ApJ...725.2038D}.
Now the shock velocities we infer is $v_{s} \sim 5,100$--$6,800$ km s$^{-1}$.
In either case, the intrinsic shock velocity of the inward shocks is some  $\sim$1--2 times higher than that of the forward shocks
\citep*[4,200--5,200 km s$^{-1}$:][]{2009ApJ...697..535P}.

The temperature of the medium (cool or hot plasma) into which the shock wave is propogating changes the conditions of the shock. Assuming first the cool
ejecta as the medium, the sound speed (a few tens km s$^{-1}$) is much lower than the shock velocity ($\sim$7,000--8,700 km s$^{-1}$) implying a high
Mach number $\mathcal{M} \gtrsim 100$. In this case, the shock compression ratio is $r = 4$. On the other hand, the Mach number of the shock could be much lower,
if the inward shock is propagating into hot plasma with a high sound speed. Using a typical value for the  electron temperature for Cassiopeia
A \citep[2 keV, e.g.,][]{2012ApJ...746..130H}, the sound speed $c_s$ and Mach number $\mathcal{M}$ are estimated to be $c_s\sim$ 730 km s$^{-1}$ and $\mathcal{M} \sim$
7--8 (for $v_{s} \sim 5,100$--$6,800$ km s$^{-1}$), Then the compression ratio $r = \frac{(\gamma+1)\mathcal{M}^2}{(\gamma-1)\mathcal{M}^2+2}$ is estimated to
be $\sim$3.8 for $\gamma = 5/3$ and $\mathcal{M} =$ 7--8, which is close to the high Mach number limit of $r = 4$. In contrast, the forward shock propagates
though the ISM, where both a high Mach number and high compression ratio are generally expected. We here note that effects of particle acceleration
can produce a shock with a compression ratio greater $r = 4$ \citep[e.g.,][]{1999ApJ...526..385B}. As relativistic particles are produced and
contribute significantly to the total pressure, the shocked plasma becomes more compressible ($\gamma \to$ 4/3, $r \to$ 7). Additionally the escape
of the highest energy particles would also further compresses the shock by carrying off energy, analogous to that case of radiative shocks
\citep[e.g.,][]{1984ApJ...277..429E,1999ApJ...526..385B,2005A&A...429..569E}.

A further complication for the shocked plasma in the remnant concerns the possibility that the electron and ion temperatures are not equilibrated. In the  extreme case, the ions
possess virually all of the thermal energy. \cite{2001ApJ...563..828L} have calculated the time variation of the ion and electron temperatures for reverse shocks
into pure oxygen ejecta for Cassiopeia A. Based on his results, we assume a plasma temperature of $\sim$ 46 keV, which is the ion temperature in the shocked plasma 150 yr after
explosion\footnote{See Table 3 and Fig.3 in \cite{2001ApJ...563..828L} for more details. The plasma parameters in this case explained well typical values
in the remnant (e.g., $k_{\rm e}T \sim$ 2.3 keV, $n_{\rm e}t \sim$ 7$\times$10$^{10}$ cm$^{3}$ s$^{-1}$).}. Here the sound speed $c_s$ and the Mach number $\mathcal{M}$ are
estimated to be $c_s\sim$ 3,500 km s$^{-1}$ and $\mathcal{M} \sim$ 1.5--1.9. Also, the compression ratio is estimated to be $\sim$2 for $\gamma = 5/3$ and $\mathcal{M} =$ 1.7.
However, it is possible that particles would not accelerated in this case because the Mach number is below the critical value of $\mathcal{M} < \sqrt{5}$ for particle acceleration
\citep{2014ApJ...780..125V}.  Columns 1 and 2  of Table \ref{tab:para}  summarize the  compression ratios and the shock velocities  for the several cases we study here.
Given the current state of knowledge,  all of these cases offer plausible scenarios for the forward and inward shocks.

Determining the shock conditions more accurately requires measuring the ion temperatures via, e.g.,  thermal line broadening in the X-ray spectra.
From {\it Chandra} grating spectroscopy of some bright knots \citep[e.g.,][]{2006ApJ...651..250L}, ion temperatures were estimated
to be quite low (the observed lines were narrow). Therefore, the high Mach number scenario may be realistic at least for some dense features.
%On the other hand, it is difficult to measure with the current X-ray detectors.
%In particular, the spectra of Cassiopeia A have a complex Doppler broadening at the same time.
%For an accurate estimation of the broadenings,
On the other hand, for diffuse extended sources like Cassiopeia A, it is difficult to measure the line broadening accurately using grating spectroscopy.
In the future, observations with a X-ray calorimeter would be more powerful. For example, the high energy resolution (FWHM $\simeq$ 5 eV) calorimeter on-board the {\it Hitomi} satellite
measured line broadening of $\sim$200 km s$^{-1}$ in the Perseus cluster \citep{2016Natur.535..117H,2017arXiv171100240H} and would have revealed much about the
thermal broadening in  SNRs, too.
We await the launch of {\it XARM}, the recovery mission for the {\it Hitomi} satellite, to reveal the shock condition more accurately
in Cassiopeia A.

\subsection{The Diffusion Coefficient and Particle Acceleration in Cassiopeia A} \label{sec:dis2}
\begin{table*}[t]
\caption{
Summary of diffusion and acceleration parameters estimated for the forward shock and the reverse shock\tablenotemark{1}.
}
\begin{center}
\begin{tabular}{lllllllll}
\hline
&	 compression ratio 						& 		shock velocity 							&		$E_{\gamma,{\rm cut},{\rm keV}}$	& 		$B$								&	$k_0$ 			&	$E_{\rm e,max}$	&	$\tau_{\rm syn}$		&	$\tau_{\rm acc}$	\\ \hline
\multicolumn{2}{l}{\bf forward shock} 		&												&											&										&					&					&							&						\\
&		$r = 4$ 							& 		4,200--5,200 km s$^{-1~(2)}$			&			2.3 keV$^{(1)}$					& 	0.1 mG$^{(2),(3)}$					&	1.1--1.6		&	34 TeV			&			37 yr			&		63 yr			\\
&		$\prime\prime$ 						& 		$\prime\prime$							&			$\prime\prime$					& 	0.5 mG$^{(4)}$						&	$\prime\prime$	&	15 TeV			&			3 yr			&		6 yr			\\
&		$r = 7$ 							& 		$\prime\prime$							&			$\prime\prime$					& 	0.1 mG$^{(2),(3)}$					&	0.7--1.1		&	34 TeV			&			37 yr			&		63 yr			\\
&		$\prime\prime$ 						& 		$\prime\prime$							&			$\prime\prime$					& 	0.5 mG$^{(4)}$						&	$\prime\prime$	&	15 TeV			&			3 yr			&		6 yr			\\
\multicolumn{2}{l}{\bf inward shock} 		&												&											&										&					&					&							&						\\
&		$r = 4$ (into cool ejecta)			& 	7,000--8,700 km s$^{-1~{\rm (this~work)}}$	&			1.3 keV$^{(1)}$					& 	0.5 mG$^{\rm (5),(this~work)}$		&	5.3--8.2 		&	11 TeV			&			4 yr			&		7 yr			\\
&		$\prime\prime$						& 		$\prime\prime$							&			$\prime\prime$					& 	1.0 mG$^{\rm (5),(this~work)}$		&	$\prime\prime$ 	&	8 TeV			&			2 yr			&		3 yr			\\
&		$r = 3.8$ (into 2 keV plasma)		& 	5,100--6,800 km s$^{-1~{\rm (this~work)}}$	&			$\prime\prime$					& 	0.5 mG$^{\rm (5),(this~work)}$		&	2.9--5.1 		&	11 TeV			&			4 yr			&		7 yr			\\
&		$\prime\prime$						& 		$\prime\prime$							&			$\prime\prime$					& 	1.0 mG$^{\rm (5),(this~work)}$		&	$\prime\prime$	&	8 TeV			&			2 yr			&		3 yr			\\
&		$r = 2$ (into 46 keV plasma)		& 		$\prime\prime$							&			$\prime\prime$					& 	0.5 mG$^{\rm (5),(this~work)}$		&	3.7--6.6 		&	11 TeV			&			4 yr			&		7 yr			\\
&		$\prime\prime$						& 		$\prime\prime$							&			$\prime\prime$					& 	1.0 mG$^{\rm (5),(this~work)}$		&	$\prime\prime$	&	8 TeV			&			2 yr			&		3 yr			\\ \hline
\end{tabular}
\label{tab:para}
\end{center}
\tablenotetext{1}{The numerical values are taken from: $^{(1)}$\cite{2015ApJ...802...15G}, $^{(2)}$\cite{2009ApJ...697..535P}, $^{(3)}$\cite{2003ApJ...584..758V}, $^{(4)}$\cite{2004A&A...419L..27B}, $^{(5)}$\cite{2008ApJ...677L.105U}.}
\end{table*}

In DSA, particles are accelerated by scattering multiple times across the shock front.
Particle diffusion is characterized by the level of magnetic field turbulence. If the strength of the turbulent field is close to the unperturbed magnetic field
strength (Bohm regime: $\delta B \sim B$), particles can be efficiently scattered and accelerated \citep[see a review:][]{2008ARA&A..46...89R}. This situation is often assumed
for particle acceleration in SNRs, although it might not be so in some cases \citep*[e.g.,][]{2006A&A...453..387P,2011ApJ...728L..28E}.
In the following we follow \cite{2006A&A...453..387P} who characterize departures from the Bohm limit  as $k_0 = D_0/D_{\rm 0,Bohm}$, where $D_0$ is the diffusion
coefficient at the electron cut-off energy. This parameter can be expressed in terms of the photon cut-off energy ($E_{\gamma,cut} = h\nu_{\rm cut}$),
shock velocity ($V_3$, in units of 1000 km s$^{-1}$) and the compression ratio $r$ as
\begin{equation}
k_0 = 0.14 ~ E^{-1}_{\gamma,{\rm cut},{\rm keV}} ~ V^2_{3,{\rm shock}} ~ \frac{16(r-1)}{3r^2}. \label{k0}
\end{equation}
The {\it NuSTAR} spectra above 15 keV showed  photon cut-off energies for the forward shock and the reverse shock of 2.3 keV and 1.3 keV, respectively \citep{2015ApJ...802...15G}.
We here assume the shock velocity and the compression ratio at the forward shock and the inward shock listed in Table \ref{tab:para}, which are estimated in the section \ref{sec:dis1}.
For the forward shock, the diffusion parameter was $k_0 \lesssim$ 1.6, which is close to the Bohm regime, in both of the compression ratios ($r = 4, 7$).
On the other hand, the inward shock shows departures from Bohm diffusion: $k_0 \gtrsim$ 3 (Table \ref{tab:para}).
\cite{2006NatPh...2..614S} have already suggested a similar $k_0$ difference between the forward shock and the inward shock in their map of the electron cutoff frequencies
in Cassiopeia A. The authors estimated upper limits to the diffusion coefficient and found that the forward-shock in
the north, northeast, and southeast was close to the Bohm limit. These
results agree with ours.
% that the scattering frequency in the forward shock is
%very larger than in the inward shock because the diffusion is closer
%to the Bohm limit.

Using the coefficient $k_0$ and other observational values, we can also estimate the electron maximum energy $E_{\rm e,max}$, the synchrotron cooling timescale
$\tau_{\rm syn}$ and the acceleration timescale $\tau_{\rm acc}$ as shown below \citep*[][]{2006A&A...453..387P};
\begin{eqnarray}
E_{\rm e,max} \simeq (8.3 ~{\rm TeV}) \frac{4}{\sqrt{3}}\frac{\sqrt{r-1}}{r} ~ k_0^{-1/2} ~ B_{0.1}^{-1/2} ~ V_{\rm sh,3},\\
\tau_{\rm syn} \simeq (1.25\times10^3 ~{\rm yr}) ~ E_{\rm TeV}^{-1} ~ B_{0.1}^{-2},\\ \label{t_syn}
\tau_{\rm acc} \simeq (30.6 ~{\rm yr}) \frac{3r^2}{16(r-1)} ~ k_0 ~ E_{\rm TeV} ~ B_{0.1}^{-1} ~ V_{\rm sh,3}^{-2}, \label{t_acc}
\end{eqnarray}
where $B_{0.1}$ and $E_{\rm TeV}$ are the magnetic field in units of 0.1 mG and the electron energy in units of 10$^{12}$ eV, respectively.
At the forward shock region, $B =$ 0.08--0.16 or 0.5 mG have been estimated by using the width of the rim of 1$^{\prime\prime}.$5--4$^{\prime\prime}$,
and also the average magnetic field in the whole SNR was estimated to be $B >$ 0.5 mG \citep[][]{2003ApJ...584..758V,2004A&A...419L..27B}. For $B \gtrsim$ 0.5 mG,
the synchrotron cooling timescale becomes comparable to the variability timescale of $\lesssim$ 4 yr at the reverse shock region \citep{2008ApJ...677L.105U}.
Our results show almost the same timescale as that in \cite{2008ApJ...677L.105U}. From these observational results, we assumed the typical
magnetic fields at the forward shock and the reverse shock are 0.1--0.5 mG and 0.5--1 mG, respectively.
Using these parameters, we newly estimate the acceleration parameters for both the forward and inward shocks of Cassiopeia A (Table \ref{tab:para}).

We find the maximum electron energy for the inward shock, $E_{\rm e,max} =$ 8--11 TeV to be smaller
than for the forward shock ($E_{\rm e,max} =$ 15--34 TeV) even though the shock velocity at
the inward shock is higher. This is because the inward shock has larger $k_0$ and $B$ values. The variability timescales are also
estimated to be shorter at the inward shock: $\tau_{\rm syn} \sim$ 2--4 yr, $\tau_{\rm acc} \sim$ 3--7 yr (e-folding time),
which is mainly due to the strong magnetic field of 0.5--1 mG. In section \ref{sec:flux}, we found the flux variations with timescales (for
changing by $\sim$40--50 \%) of $t_{\rm dec} = 4 \pm 1$ yr and $t_{\rm in} = 9 \pm 1$ yr for the W3 (and maybe also W4) region.
These timescales are very similar to the predicted timescales for $B = 0.5$ mG. For regions W1 and W2, the timescales were estimated to
be $t_{\rm in} \sim t_{\rm dec} = 4 \pm 2$ yr,
which is also close to the time scale for a strong magnetic field of 0.5--1 mG. For regions C1 and C2, we found gradual flux declines of $\sim$20--40 \%
over $\sim$14 yr. Here, $B \sim$ 0.2 mG can well explain the flux-decline timescale. From these points, we note that differences of the variability timescales can
be explained by difference in the magnetic field strength. In addition, we note that the time variation of spatially adjoining regions are similar to
each other (see both Figure \ref{fig:diff} and \ref{fig:pro_fit}). This tendency may also imply something about the difference of the magnetic field from region
to region. For example, the radial positions of the inward shocks: $R_{\rm W1,W2} > R_{\rm W3,W4} > R_{\rm C1,C2}$ seems to be inversely related to the variability
timescales: $t_{\rm W1,W2} < t_{\rm W3,W4} < t_{\rm C1,C2}$. This tendency might suggest that  magnetic field amplification has a dependence
on the radial location of the shock within the remnant. In
contrast, the variability timescales at the forward shock are much longer than at the inward shock. \cite{2017ApJ...836..225S} have shown the time
variation of X-ray flux at the forward shock region is not significant: $0.00 \pm 0.07$ \% yr$^{-1}$. The constant flux can be explained by the balance
between acceleration and cooling even for both cases of high and low magnetic fields in Table \ref{tab:para} \citep[see also][]{2010ApJ...723..383K}.
We leave for future work investigations into specific scenarios to account for the rapid flux changes that we see in some of the inward shocks.

\begin{figure}[th!]
 \begin{center}
  \includegraphics[bb=0 0 1253 1206,width=7cm]{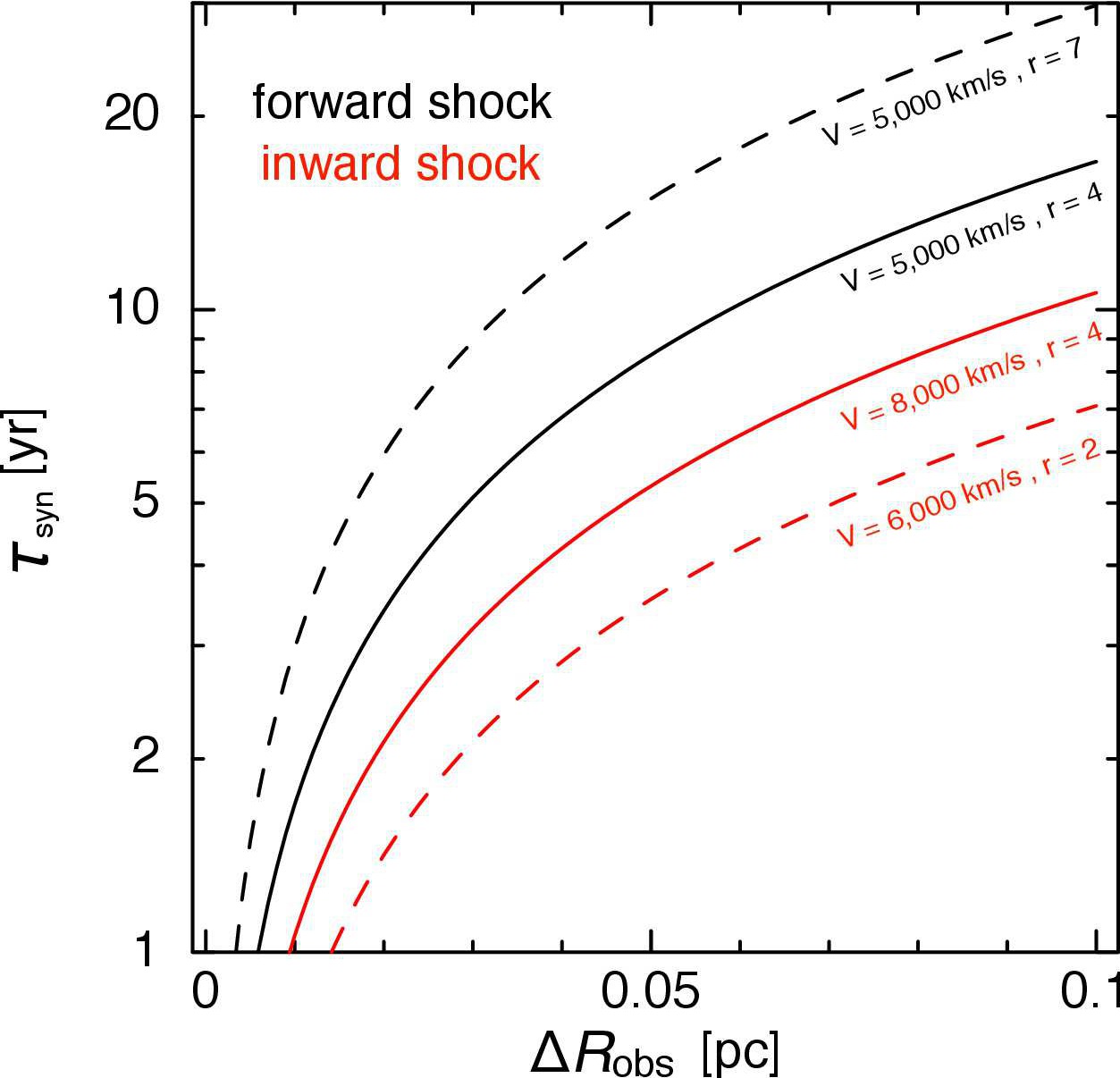}
 \end{center}
\caption{Synchrotron cooling timescale and observational shock width for the forward shock (black) and the inward shock (red) estimated from typical parameters in Table \ref{tab:para}. }
\label{fig:width}
\end{figure}

The synchrotron cooling timescale can be also estimated from the inward-shock width, $\sim$0.03--0.07 pc, presented in section \ref{sec:width}.
The accelerated electrons are advected away from the shock at a downstream velocity $V_{\rm d} = V_{\rm sh}/r$ during the synchrotron timescale,
where $r$ is the shock compression ratio. The size of the advection region is thus given by $\Delta R_{\rm adv} = V_{\rm d} \times \tau_{\rm syn}$,
where $\Delta R_{\rm adv}$ is the advection width. Although particle diffusion also expands the width, we simply constrain the observational
width as $ \Delta R_{\rm obs} \geq V_{\rm d} \times \tau_{\rm syn} \times P$, where $P$ is a projection factor. In the ideal case of a spherical
shock, the projection factor is $P = 4.6$.
\cite{2006A&A...453..387P} described the relation between the synchrotron cooling timescale and the observational width as
\begin{equation}
\tau_{\rm syn} \leq \frac{\Delta R_{\rm obs}}{V_{\rm d}} \simeq (8.5~{\rm yr}) \times \frac{r}{4} \times \bar{P}^{-1} \times V_{\rm sh,3}^{-1} \times \Delta R_{\rm obs,-2}, \label{t_width}
\end{equation}
where $\bar{P}$ and $\Delta R_{\rm obs,-2}$ are defined as $\bar{P} = P/4.6$ and $\Delta R_{\rm obs} \equiv \Delta R_{\rm obs,-2} \times 10^{-2}$ pc, respectively.
Using equation (\ref{t_width}) and the estimated parameters in Table \ref{tab:para}, the relation between the cooling time and the shock width was estimated (Figure \ref{fig:width}).
For a typical inward-shock width of 0.05 pc, the cooling timescale in the inward shock is $\lesssim$ 3--5 yr, which is consistent with the observed decay timescale and the
predicted cooling timescale in the case of $B =$ 0.5--1 mG (Table \ref{tab:para}). For the forward shock, the timescale is longer ($\sim$10 yr) than that for the inward shock.
This tendency is the same as the other observational timescale estimates (Table \ref{tab:para}).

Particle acceleration at reverse shocks have been studied in some theoretical work \citep[e.g.,][]{2005A&A...429..569E,2012A&A...541A.153T,2013A&A...552A.102T}. For all of the models,
the maximum energy reached by particles at the reverse shock is always lower than at the forward shock. Also, recent deep $\gamma$-ray observations of Cassiopeia A with MAGIC
have showed a clear exponential cut-off at $\sim$ 3.5 TeV \citep{2017arXiv170900280G}, %\citep{2014RAA....14..165Y}, which
which implies only modest cosmic-ray acceleration to very high energy for the hadronic scenario. From the observations, Cassiopeia A is considered not to be
a PeVatron (PeV accelerator) at its present age. Both theoretical and observational studies suggest that particle acceleration at the reverse shock (inward shock) in
Cassiopeia A could not produce the high energy particles (protons) up to ``knee" energy ($\sim 3 \times 10^{15}$ eV) even if the shock has as high a shock velocity as shown in
section \ref{sec:dis1}. On the other hand, the reverse shock region is propogating into a metal-rich environment produced by supernova nucleosynthesis. If such
elements were accelerated by the reverse shocks, the cosmic-ray composition would be affected. Therefore, particle acceleration at reverse shocks are important for understanding
the cosmic-ray abundances. % although the knee energy could not be anticipated.  %If although those elements might be

%\subsection{Comparison with Theoretic Views of the Reverse Shock Acceleration in SNRs} \label{sec:theory}
%The reverse shock acceleration has been predicted in some theoretical models. \cite{2005A&A...429..569E} argued that efficient accelerations at the reverse shocks with
%a non-linear effect. The non-linear effect is considered to be caused by relativistic particles \citep{1999ApJ...526..385B}. The relativistic particles are produced
%and contribute significantly to the total pressure, thus the shocked plasma becomes more compressible ($\gamma \to$ 4/3, $r \to$ 7). If the highest energy particles escape
%from the shock, they drain away energy flux. The escape effect makes also the plasma more compressible ($\gamma \to$ 1).

\subsection{The Origin of the Inward Shocks} \label{sec:dis3}
For SNR expansion into an ambient medium with a constant or gradually declining density profile (e.g., ISM, CSM), it is not possible
to explain an inward moving reverse shock (in the observer frame) for an evolutional phase that describes  Cassiopeia A. In such a young evolutionary phase,
the reverse shock must expand outward. As an alternative explanation for the inward shocks we consider a ``reflection shock''
produced by the forward shock interacting  with a density jump in the ambient medum.

\begin{figure}[h]
 \begin{center}
  \includegraphics[bb=0 0 1857 1325,width=8cm]{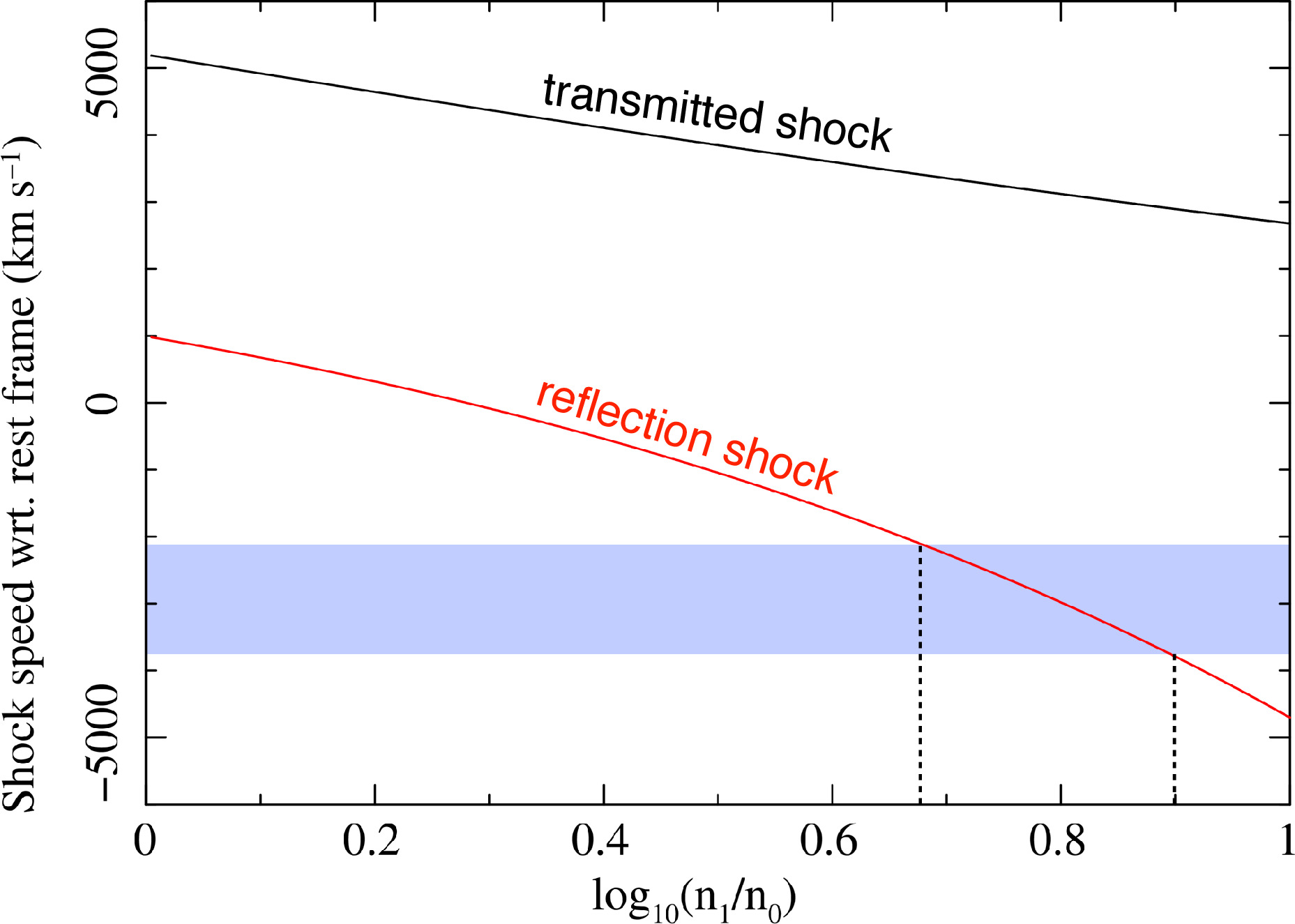}
 \end{center}
\caption{Estimate of the reflection shock velocity and the density of interacting cloud in the case of Cassiopeia A
with solving the fluid equation in \cite{1994ApJ...420..721H}. The black and red curve show the velocities of the
transmitted shock (that is propagating through the cloud) and reflection shock, respectively. Blue area shows the range of the velocities of the western filaments in this work: $\sim$2,100--3,800
km s$^{-1}$.}
\label{fig:RS}
\end{figure}

The reflection-shock conditions can be calculated using a simple fluid equation \citep{1994ApJ...420..721H}. We here assume
a density of ambient medium before the encounter with the density jump as $\rho_0$. The density jump $\rho_1$ is defined as $\rho_1 \equiv \alpha \rho_0$.
Then, the forward shock density $\rho_2$ is described as $\rho_2 = 4 \rho_0$ from the Rankine--Hugoniot relation. After the encounter, a contact discontinuity
is made by the interaction. Upward materials are compressed by $4 \rho_1$, and then downward materials are pushed back \citep[see also][]{2012ApJ...744...71I}.
Figure \ref{fig:RS} shows the estimate of the reflection shock speed using \cite{1994ApJ...420..721H}.
We assumed the forward shock is interacting with the density jump with the speed of 5,200 km s$^{-1}$.
In order to explain the inward-shock velocities, we require that the forward shock be interacting with a surrounding material whose density
is $\gtrsim$ 5-8 times higher than that of a ambient medium elsewhere.

\begin{figure}[h]
 \begin{center}
  \includegraphics[bb=0 0 557 689,width=8cm]{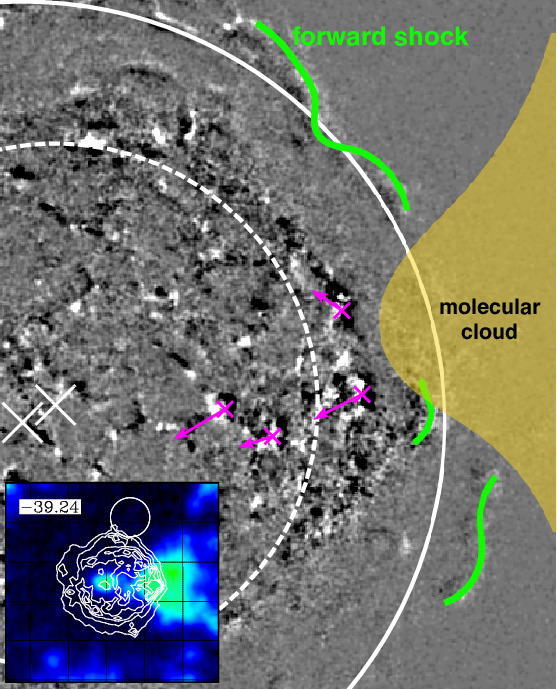}
 \end{center}
\caption{A schematic diagram of the shock-cloud interaction. The image show the same image as in Figure \ref{fig:diff}(a).
Magenta arrows show the proper-motion directions of the inward shocks. The forward shocks are emphasized by thick green lines.
Small and large circles indicate the radii of the reverse shock and forward shock \citep{2001ApJ...552L..39G}, respectively.
Cross marks show the center positions of each circle. The center of the reverse-shock circle offsets to the northeast direction.
Small screen set on left bottom show the CO map with the velocity of $-39.24$ km s$^{-1}$ in \cite{2014ApJ...796..144K}.
A distance between the reverse shock and the forward shock is $\sim$ 0.7 pc at the western region.}
\label{fig:comp}
\end{figure}

A local molecular cloud would be a good candidate for explaining such an enhanced density structure around the remnant. Millimeter observations
in $^{12}$CO and $^{13}$CO J = 2--1 (230 and 220 GHz) with the Heinrich Hertz Submillimeter Telescope indicated the existence of the molecular
cloud around Cassiopeia A \citep{2014ApJ...796..144K}. In particular, the inward-shock regions seems to overlay with the fastest gas
($-50$ to $-45$ km s$^{-1}$) and the slowest gas ($-39$ to $-34$ km s$^{-1}$). We show a schematic diagram of the shock-cloud interaction in
Figure \ref{fig:comp}. As shown in this figure, the inward shocks seems to be shifting radially against the distribution of the molecular clouds.
Here, we can estimate the time when the forward shock first hits the cloud and pushes the reflection shock into the ejecta.
%the timescale since the forward shock hits the cloud can be estimated to be $<$ 100 yrs,
Dividing the distance between the forward shock and the inward shocks ($\sim$ 0.7 pc, see Figure \ref{fig:comp}) by their velocity difference
of $\sim$ 7000 km s$^{-1}$, it can be estimated to be $<$ 100 yrs ago. This is much shorter than the age of the remnant ($\sim$ 350 yrs old),
providing us with a further support that the inward shocks are reflection shocks
by the western cloud. Not only the western filaments but also the filaments at the central position (e.g., C1, C2) might be also related to the
shock-cloud interaction. For example, the high-speed CO gas ($-46$ to $-49$ km s$^{-1}$) is concentrated in filamentary structure to the south
and southeast of the remnant. The locations are very close to the positions of the central inward shocks \citep[see Figure 3 right in][]{2014ApJ...796..144K}.

However, it is difficult to confirm the shock-cloud interaction. \cite{2014ApJ...796..144K} have argued for a shock-cloud interaction
around the western and southern rim using the broadened CO lines \citep[see Fig. 4 in][]{2014ApJ...796..144K}. As a matter of fact, their locations
are a little different from the inward-shock positions. Future observations
with the Nobeyama 45-m Telescope will be helpful in studying the shock interaction with the molecular cloud.  $^{12}$CO observations
with the telescope would have good spatial resolution ($\sim 17^{\prime\prime}$) able to reveal more small structures and the relation with the X-ray
distributions (Inaba et al., private communication). For example, in the case of RX J1713.7--3946, non-thermal X-rays which are enhanced around
CO and H$_{\rm I}$ clumps have been found \citep[e.g.,][]{2010ApJ...724...59S,2013ApJ...778...59S,2015ApJ...799..175S}.
They suggested that the amplified magnetic field around the CO and H$_{\rm I}$ clumps enhances the synchrotron X-rays and possibly the
acceleration of cosmic-ray electrons. In the theoretical view, the amplified magnetic field could be explained as one of features
of the shock-cloud interaction. \cite{2012ApJ...744...71I} have investigated cosmic-ray acceleration assuming interaction with
clumpy interstellar clouds using three-dimensional magnetohydrodynamic simulations. They predicted a highly amplified magnetic field of
$\sim$1 mG caused by a turbulent shell due to the shock-cloud interactions. Then, the short-time X-ray variability was predicted at the same time.
This supports well our observational results in Cassiopeia A as discussed in the section \ref{sec:dis2}.

%If the shocks of Cassiopeia A are actually interacting with the molecular clouds, the high amplified magnetic field is also explained as one of features
%of the shock-cloud interaction.

%In this simulation, the existence of the reflected shocks was
%also suggested. The Mach number of the reflection shock was estimated to be $\sim$1.8 (the maximum limit is $\sim$2.24), which is almost consistent with
%our estimations assuming the hot plasma (40 keV).
%Therefore, the cosmic-ray accelerations at the reflection shocks were not predicted. On the other hand, they assumed a plane
%parallel shock. Although the physical solution for the parallel shock is $M \sim$ 2.24 for the reflection shock, the real SNR shock could be over this value
%(Inoue Tsuyoshi in private communication). Actually, our observational Mach number indicated the strong shock. Therefore, the rapid acceleration and cooling at
%the reflection shocks caused by the shock-cloud interaction would be able to happen.

\section{Conclusion} \label{sec:conclusion}
The bright non-thermal X-ray emission in the interior of Cassiopeia A
has been one of the most enigmatic features of the remnant since the
earliest  observations by  {\it Chandra}.
%
%One of mysterious features in Cassiopeia A is the bright non-thermal X-ray emission at the interior
%of the remnant.
%
Even as basic a fact as the type of shock (e.g., forward shock,
reverse shock or something else) has remained obscure.  In this paper,
we put forth new evidence that the interior non-thermal emission
originates in an ``inward shock'' through new analyses of archival
{\it Chandra} and {\it NuSTAR} observations. We identified inward
moving filaments in the remnant's interior using monitoring data by
{\it Chandra} from 2000 to 2014. The inward-shock positions are spatially
coincident with the most intense hard (15--40 keV band) X-ray emission seen
with {\it NuSTAR}.
%This result suggests the hard X-rays are emitted
%mainly at the reverse shock (=inward shock).

We measured the proper motions of the inward shocks, which equate to
speeds of $\sim$2,100--3,800 km s$^{-1}$ for a distance of 3.4 kpc to
the remnant.  Assuming the shocks are propagating through the
expanding ejecta (which is itself moving outward), we determine that
the shock velocities in the frame of the ejecta could reach up to
$\sim$5,100--8,700 km s$^{-1}$, which is $\sim$ 1--2 times higher than
that at the forward shock.  Additionally some of the inward-shock
filaments showed flux variations (both increasing and decreasing) on
timescales of just a few years. We find that the high shock velocity
combined with a high magnetic field strength ($\sim$ 0.5--1 mG) in the
reverse shock region can explain the non-thermal properties well. At
the same time we are able to constrain the diffusion coefficient and
find that diffusion at the reverse shock is less efficient ($k_0
\gtrsim$3) than that of the forward shock ($k_0 \lesssim$ 1.6).
Expressed in terms of magnetic field turbulence, we find (as do
others) that the turbulence at the forward shock approaches the Bohm
limit, while at the inward shocks turbulence is less well developed.

As to the nature of the inward shocks, we propose that they are
``reflection shocks'' caused by the forward shock's interaction with a
density enhancement in the circumstellar medium. A density jump of a
factor of $\gtrsim$5--8 reproduces the observed inward-shock
velocities.  Previous works have shown evidence for a local molecular
cloud on the western side of Cassiopeia A with some indications of an
interaction beyween the cloud and the remnant's shock.  Further
investigations into the shock-cloud interaction will be useful to
deepen our understanding of particle acceleration in Cassiopeia A.

\acknowledgments
We thank the {\it Chandra} and {\it NuSTAR} Operations, Software, and Calibration teams for support
with the execution and analysis of these excellent observations.
This work was supported by the Japan Society for the Promotion of Science (JSPS) KAKENHI Grant Number
16J03448 (T.S), 16K17673 (S.K) and 17K05395 (M.M). Also, M.M is supported by CREST. F.F was supported,
in part, by NASA under Grant NNX15AJ71G. This research was also supported in part by NASA grant NNX15AK71G
to Rutgers University.  J.P.H.\ acknowledges the hospitality of the Flatiron Institute which is supported
by the Simons Foundation. We thank Tetsuta Inaba, Hidetoshi Sano, Yasuo Fukui, Tsuyoshi Inoue, Ryo Iizuka,
Takaaki Tanaka, Hiroyuki Uchida, Kuniaki Masai, Shiro Ikeda, Robert A. Fesen and Daniel J. Patnaude
for helpful discussions and suggestions in preparing this paper. We also thank the anonymous
referee for his/her comments that helped us to improve the manuscript.
%M.M is supported by the Japan Society for the Promotion of Science (JSPS) KAKENHI grant number 17K05395.
 %K.S is supported by the Japan Society for the Promotion of Science
%#(JSPS) KAKENHI grant number 16K17673.

 \software{NuSTARDAS (v1.6.0), CIAO \citep[version 4.6;][]{2006SPIE.6270E..1VF},
OpenCV (v3.2.0; [http://opencv.org]), XSPEC \citep[v12.9.0n;][]{1996ASPC..101...17A})}

\appendix
\section{Rice Distribution} \label{sec:rice}

In our study we determined the magnitude of proper motion $\mu$
and its direction $\theta$ from measurements along the RA and
decl.\ axes.  To sufficient accuracy we can approximate our
measurements as normally distributed in $X$ and $Y$ with means of
$\mu \cos(\theta)$ and $\mu \sin(\theta)$ and variances of $\sigma^2$.
 This formulation assumes that the uncertainties are
the same in both directions. Our naive estimate for the proper
motion, $p =(X^2 + Y^2)^{1/2}$, follows the Rice distribution
\citep[see also Appendix I in][]{1958AcA.....8..135S}, which
has a probability density function (PDF) given by

\begin{equation}
 f(p) =  {p \over \sigma^2} \exp \left ( -\frac{p^2 + \mu^2}{2\sigma^2} \right )
        I_0 \left(\frac{p\mu}{\sigma^2} \right)
\end{equation}
where $I_0$ is the modified Bessel function of the first kind with
order zero.  This distribution function is bounded at zero and is
asymmetric, especially when $\mu/\sigma^2$ is small.  Hence the naive
estimate for the magnitude of proper motion just introduced is biased
high.  In order to account for this bias, we use the Rice PDF.

The mean of the Rice PDF is

\begin{equation}
 \langle \mu \rangle =  \sigma \sqrt{\pi/2}\, \exp(-\mu^2/2\sigma^2)\,
_1 F_1 (3/2;1;\mu^2/2\sigma^2) , \label{A2}
\end{equation}
\noindent
where $_1 F_1$ is the confluent hypergeometric function.  This can be
expressed in terms of Bessel functions \citep[e.g.,][]{1991ASAJ...89.1193T} as

\begin{equation}
 \langle \mu \rangle =  \sigma \sqrt{\pi/2}\, e^{-x/2}\,
\left[ (1+x) I_0(x/2) +x I_1(x/2)\right]  \label{A3}
\end{equation}
\noindent
with $x=\mu^2/2\sigma^2$.  The variance of the Rice distribution is given by

\begin{equation}
 \sigma^2_{\langle \mu  \rangle} =  2\sigma^2 + \mu^2 - \langle \mu \rangle^2 . \label{A4}
\end{equation}

To obtain an unbiased estimate for the proper notion we solve Eq. (\ref{A3})
for $\mu$ by setting $p= \langle \mu \rangle$.  Eq. (\ref{A4}) is then used
to determine the uncertainty on the proper motion.

\listofchanges

\end{document}